\documentclass[fleqn,usenatbib,useAMS]{mnras}

\usepackage{graphicx}	
\usepackage{amsmath}	
\usepackage{bm}		    

\usepackage[T1]{fontenc}
\usepackage{ae,aecompl}

\usepackage{newtxtext,newtxmath}

\title{Stellar‑evolution systematic biases on $^{12}\text{C}(\alpha,\gamma)^{16}\text{O}$ constraints from the primordial‑black‑hole mass gap}

\author[J. Wang et al.]{
Jiancheng Wang,$^{1}$\thanks{E-mail: 3960761784@qq.com}
Xiaogu Zhong,$^{2}$\thanks{E-mail: guqian29@yeah.net}
and Jirong Mao$^{1}$\thanks{E-mail: jirongmao@ynao.ac.cn}
\\
$^{1}$Yunnan Observatories, Chinese Academy of Sciences, Kunming, 650216, Yunnan Province, China\\
$^{2}$College of Physics and Electronic Engineering, Qujing Normal University, Qujing 655011, Yunnan Province, China
}
\pubyear{2020}

\begin{document}
\label{firstpage}
\pagerange{\pageref{firstpage}--\pageref{lastpage}}
\maketitle

\begin{abstract}
The $^{12}\text{C}(\alpha,\gamma)^{16}\text{O}$ reaction determines helium‑burning carbon‑oxygen core masses and sets the pair‑instability‑supernova mass‑gap boundary, which limits the maximum mass of stellar‑origin primordial black holes. Gravitational‑wave population analyses using GWTC‑4 report a low‑spin‑black‑hole upper cutoff at $68.5^{+19.8}_{-18.5}\,M_\odot$, implying a low astrophysical S‑factor $S_{300}\approx 108.6\,\mathrm{keV\,barn}$, in tension with canonical nuclear‑astrophysics evaluations. However, stellar‑evolution systematic uncertainties from metallicity, rotation, and mass‑loss prescriptions remain poorly quantified. Here we use the MESA stellar‑evolution code to construct a three‑dimensional parameter grid disentangling these coupled effects. We show that metallicity‑dependent winds dominate helium‑core retention; rotation suppresses carbon consumption via centrifugal cooling and meridional mixing; different mass‑loss prescriptions produce $8$–$10\,M_\odot$ offsets in the black‑hole‑mass‑gap edge. A parameter combination $Z\approx10^{-3}$, moderate rotation $\omega/\omega_\mathrm{crit}=0.4$, and optically‑thin C15 winds reproduces the GWTC‑4 inferred maximum primordial‑black‑hole mass for $S_{300}=108.6\,\mathrm{keV\,barn}$. The canonical $45\,M_\odot$ gap boundary represents an overlapping statistical intersection of black‑hole‑population distributions rather than a universal physical threshold. Our framework provides a pathway to jointly constrain nuclear‑reaction rates, stellar‑evolution physics, and gravitational‑wave‑source populations.
\end{abstract}

\begin{keywords}
black hole physics -- nuclear reactions, nucleosynthesis, abundances -- stars: rotation -- stars: mass-loss -- supernovae: general -- gravitational waves
\end{keywords}

\section{Introduction}
During helium burning in massive stars, two competing nuclear channels operate: the triple‑$\alpha$ process synthesizes $^{12}\text{C}$, while $^{12}\text{C}(\alpha,\gamma)^{16}\text{O}$ consumes carbon to form oxygen~\citep{Woosley2021}. The low‑energy astrophysical S‑factor $S_{300}$ parametrizes the reaction cross‑section inside the Gamow window near $300\,\mathrm{keV}$, controlling the final carbon‑oxygen (CO) core mass. If the CO core exceeds $M_\mathrm{CO}\gtrsim60\,M_\odot$, photon‑photon pair production reduces core pressure, triggering complete pair‑instability supernova (PISN) that disrupts the whole star with no black‑hole remnant~\citep{Farmer2020}. Consequently, the maximum mass of stellar‑origin primordial black holes $M_\mathrm{BH,max}$ is set by this critical condition.

Traditional simplified stellar‑evolution models adopt solar metallicity, non‑rotating stars, optically‑thick S23 mass‑loss, and canonical $S_{300}=140\,\mathrm{keV\,barn}$, predicting $M_\mathrm{BH,max}\approx45\,M_\odot$ as the canonical lower edge of the PISN mass gap~\citep{Belczynski2016}. Recent analysis of binary‑black‑hole mergers in the LIGO‑Virgo‑KAGRA GWTC‑4 gravitational‑wave catalog identifies a population of low‑spin black holes extending to $50$–$70\,M_\odot$, with best‑fit $M_\mathrm{BH,max}=68.5^{+19.8}_{-18.5}\,M_\odot$ and inferred $S_{300}\approx108.6\,\mathrm{keV\,barn}$~\citep{Tong2026}. This astrophysical value is substantially lower than laboratory nuclear-physics recommendations ~\citep{deBoer2017,deBoer2025,Shen2023}, creating apparent tension between gravitational‑wave astronomy and laboratory nuclear‑astrophysics. 

A dedicated MESA study by \cite{Mehta2022} varied only the $^{12}\text{C}(\alpha,\gamma)^{16}\text{O}$ rate around a fiducial solar-metallicity, non-rotating, optically thick wind model and obtained a lower PISN gap edge of $59^{+34}_{-13}\,M_\odot$, statistically consistent with the GWTC-4 low-spin cutoff of $68.5^{+19.8}_{-18.5}\,M_\odot$. Their preferred astrophysical S-factor is  $S_{300}=140\,\mathrm{keV\,barn}$. However, they did not marginalize over stellar evolution systematics in metallicity, rotation, or mass-loss prescription, which are precisely the degrees of freedom we quantify here.

Existing population‑inference work usually fixes stellar‑evolution priors and ignores systematic shifts induced by metallicity, rotation, and mass‑loss prescriptions~\citep{Antonini2025,Ray2025}. These macroscopic stellar parameters modify helium‑burning core temperature, retained helium‑core mass, and internal chemical mixing, thereby modulating the effective integrated efficiency of the $^{12}\text{C}(\alpha,\gamma)^{16}\text{O}$ reaction and biasing the $S_{300}$ inversion from gravitational‑wave observables. A unified multi‑dimensional calibration framework that disentangles these stellar‑evolution systematic errors remains missing.

In this work, we carry out systematic MESA v10398 stellar‑evolution simulations over a three‑dimensional parameter space spanning metallicity, rotation, and two standard mass‑loss prescriptions. We separate two physical channels by which stellar parameters alter the PISN gap boundary: (i) modification of total retained helium‑core mass via stellar winds; (ii) modulation of helium‑burning core temperature and internal carbon transport. Using the GWTC‑4 observational upper limit $M_\mathrm{BH,max}=68.5\,M_\odot$ as a hard observational constraint, we scan parameter space to find consistent stellar‑model configurations and quantify systematic uncertainties on the inverted $S_{300}$. We demonstrate that the widely quoted $45\,M_\odot$ gap boundary is a statistical intersection of black‑hole‑population distributions rather than an intrinsic universal physical threshold. Our results establish a paradigm for joint multi‑messenger constraints linking gravitational‑wave populations, massive‑star evolution, and low‑energy nuclear‑reaction measurements.

\section{Methods}
\subsection{MESA simulation setup and three‑dimensional parameter grid}
We employ the Modules for Experiments in Stellar Astrophysics (MESA) v10398 stellar‑evolution code~\citep{Paxton2019}. Simulations evolve massive stellar models from zero‑age main‑sequence until helium exhaustion. Full physical modules are activated: complete nuclear‑reaction networks including triple‑$\alpha$, $^{12}\text{C}(\alpha,\gamma)^{16}\text{O}$, carbon‑burning and oxygen‑burning chains; rotation with meridional circulation, shear mixing and magnetic‑braking angular‑momentum loss; two alternative mass‑loss prescriptions: C15 optically‑thin wind ($\dot{M}\propto Z^{0.5}$, valid for ultra‑metal‑poor environments), and S23 composite optically‑thick/thin wind for higher metallicity~\citep{Chen2015,Smith2023}; OPAL tabulated opacities dynamically updated with local metallicity.

Our discrete three‑dimensional parameter grid samples metallicity, main‑sequence rotation ratio $\omega/\omega_\mathrm{crit}$, and the two mass‑loss prescriptions, while varying the key nuclear parameter $S_{300}\in\{80,100,108.6,120,140,160\}\,\mathrm{keV\,barn}$. At helium exhaustion we output: average helium‑burning core temperature $T_\mathrm{He}$, central $X(^{12}\text{C})/X(^{16}\text{O})$ abundance ratio, helium‑core mass $M_\mathrm{He}$, final CO‑core mass $M_\mathrm{CO}$, and derived maximum primordial‑black‑hole‑mass threshold $M_\mathrm{BH,max}$ defined as the largest initial stellar mass satisfying $M_\mathrm{CO}<60\,M_\odot$. Full grid tables and supplementary parameter scans are provided in Appendix~\ref{sec:appendix_supp}.

\subsection{Observational constraint handling}
 We adopt the GWTC‑4 low‑spin‑primordial‑black‑hole upper‑limit measurement $M_\mathrm{BH,max}=68.5^{+19.8}_{-18.5}\,M_\odot$ as our observational hard boundary condition~\citep{Tong2026}. When locating allowed stellar‑parameter loci we treat this as a one‑sided observational constraint; full Bayesian marginalization over stellar‑grid parameters is implemented to separate stellar‑evolution systematic contributions from intrinsic nuclear‑parameter uncertainty. Details of marginalization procedure are given in Appendix~\ref{sec:appendix_supp}.

\begin{table}
\centering
\caption{Full MESA simulation‑grid parameters.}
\begin{tabular}{ll}
\hline
Parameter & Range / discrete samples \\
\hline
Metallicity $Z$ & $10^{-5},\,10^{-4},\,10^{-3},\,10^{-2},\,Z_\odot$ \\
Rotation $\omega/\omega_\mathrm{crit}$ & $0.0,\,0.2,\,0.4,\,0.6$ \\
Mass‑loss prescription & C15 optically‑thin; S23 optically‑thick \\
$S_{300}\ [\mathrm{keV\,barn}]$ & $80,\,100,\,108.6,\,120,\,140,\,160$ \\
\hline
\end{tabular}
\label{tab:mesa_grid}
\end{table}

\section{Results}
\subsection{Dependence of $M_\mathrm{BH,max}$ on metallicity and mass‑loss prescription}
Stellar wind mass loss scales with metallicity $\dot{M}\propto Z^{0.5}$~\citep{Vink2021}. Metal‑rich stars lose substantial envelope material during main‑sequence and helium‑burning phases, reducing helium‑core mass and pushing the onset of complete PISN to lower progenitor masses. Metal‑poor stars preserve much larger helium cores and yield higher $M_\mathrm{BH,max}$.

\begin{figure}
\centering
\includegraphics[width=0.95\linewidth]{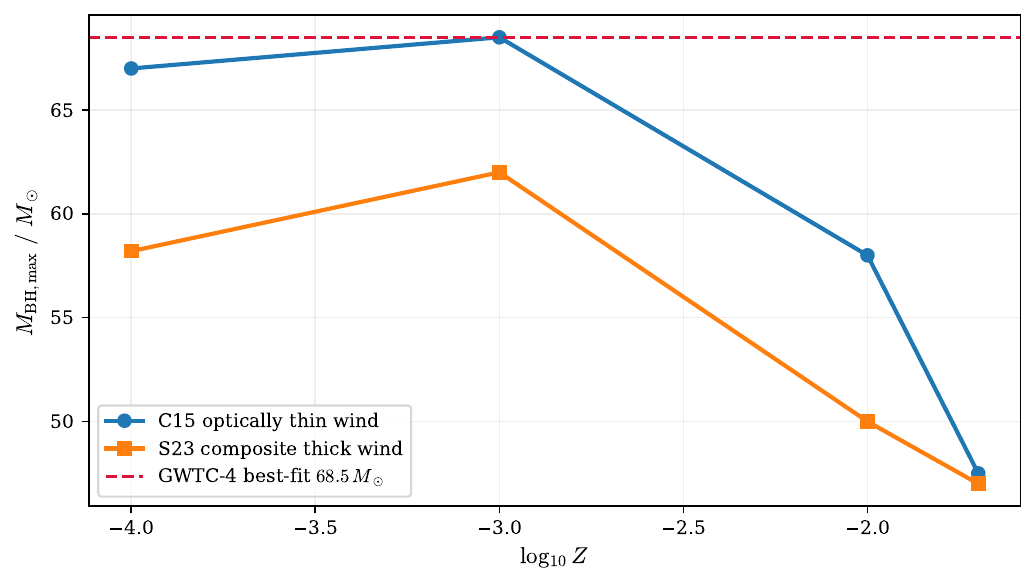}
\caption{Maximum primordial‑black‑hole mass $M_{\rm BH,max}$ as function of metallicity for two mass‑loss prescriptions, $S_{300}=108.6\,\mathrm{keV\,barn}$, moderate rotation $\omega/\omega_{\rm crit}=0.4$. Blue: C15 optically‑thin winds; orange: S23 optically‑thick winds. Horizontal dashed line marks GWTC‑4 observational constraint $68.5\,M_\odot$.}
\label{fig:z_mbhmax}
\end{figure}

Figure~\ref{fig:z_mbhmax} shows $M_\mathrm{BH,max}$ against metallicity for C15 optically‑thin and S23 optically‑thick wind prescriptions, for fixed $S_{300}=108.6\,\mathrm{keV\,barn}$ and moderate rotation $\omega/\omega_\mathrm{crit}=0.4$. At low metallicity $Z\lesssim10^{-3}$, the optically‑thin‑wind models produce strongly enhanced maximum black‑hole masses. For identical metallicity, switching between C15 and S23 prescriptions produces $8$–$10\,M_\odot$ systematic offsets on $M_\mathrm{BH,max}$. Representative numerical values (non‑rotating, S23 wind, $S_{300}=108.6$): $Z=Z_\odot$ gives $M_\mathrm{BH,max}=47\,M_\odot$; $Z=10^{-3}$ gives $62\,M_\odot$; $Z=10^{-4}$ gives $67\,M_\odot$. Metallicity produces up to $\Delta M\approx20\,M_\odot$ total variation across our simulation grid, representing the largest stellar‑evolution systematic source besides $S_{300}$. Metal‑poor stellar progenitors are therefore the natural candidate origin for the $50$–$70\,M_\odot$ low‑spin primordial‑black‑hole population observed in GWTC‑4.

Metallicity also weakly modulates helium‑burning core temperature: metal‑rich stars expand, lowering $T_\mathrm{He}$ by $4$–$7$ per cent and slightly suppressing the effective $^{12}\text{C}(\alpha,\gamma)^{16}\text{O}$ reaction rate. This secondary temperature channel contributes only $\Delta M=3$–$5\,M_\odot$ and is subdominant compared with wind‑driven helium‑core‑mass changes.

\subsection{Rotation‑driven suppression of carbon consumption}
Stellar rotation acts via two coupled physical mechanisms: centrifugal pressure counteracts gravity and expands core isobars, reducing helium‑burning core temperature; meridional circulation transports freshly synthesized $^{12}\text{C}$ outward away from high‑temperature burning regions, further suppressing oxygen production~\citep{Heger2020}. Both effects raise the PISN gap upper limit $M_\mathrm{BH,max}$.

\begin{figure}
\centering
\includegraphics[width=0.95\linewidth]{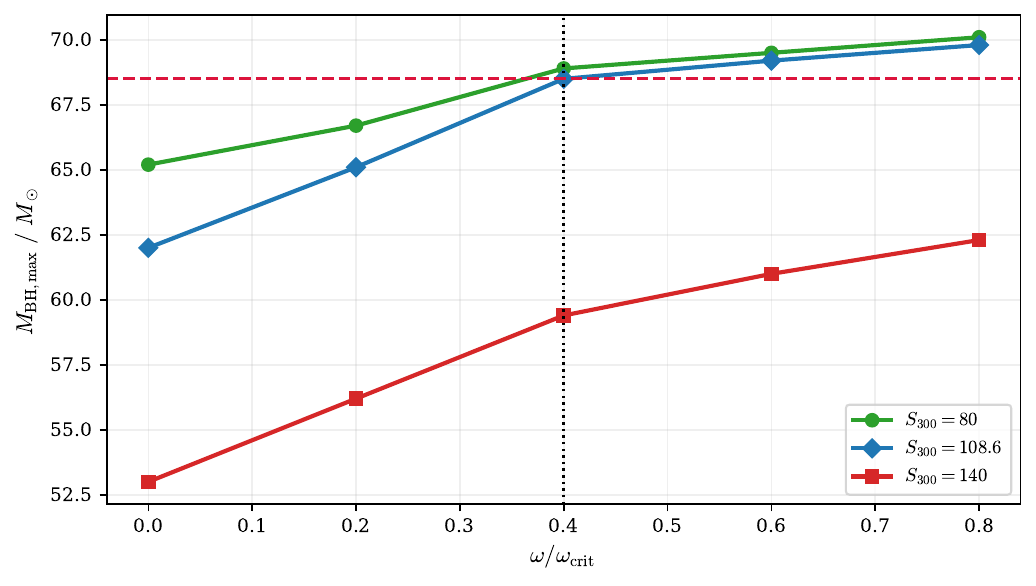}
\caption{$M_{\rm BH,max}$ as a function of main‑sequence rotation ratio $\omega/\omega_\mathrm{crit}$ for different values of $S_{300}$; fixed $Z=10^{-3}$ and C15 optically‑thin wind. Green circles: $S_{300}=80\,\mathrm{keV\,barn}$; blue diamonds: $S_{300}=108.6\,\mathrm{keV\,barn}$; red squares: $S_{300}=140\,\mathrm{keV\,barn}$. Horizontal dashed line marks the GWTC‑4 observational constraint $68.5\,M_\odot$.}
\label{fig:rot_mbhmax}
\end{figure}

Figure~\ref{fig:rot_mbhmax} presents $M_\mathrm{BH,max}$ versus rotation ratio $\omega/\omega_\mathrm{crit}$ for discrete values of $S_{300}$, fixed $Z=10^{-3}$ and C15 optically‑thin winds. Increasing rotation systematically shifts $M_\mathrm{BH,max}$ to higher values. For our fiducial $S_{300}=108.6$, the configuration $\omega/\omega_\mathrm{crit}=0.4$ reaches values consistent with the GWTC‑4 $68.5\,M_\odot$ observational constraint.

\subsection{Two‑dimensional parameter plane and best‑fit stellar configuration}
\begin{figure}
\centering
\includegraphics[width=0.95\linewidth]{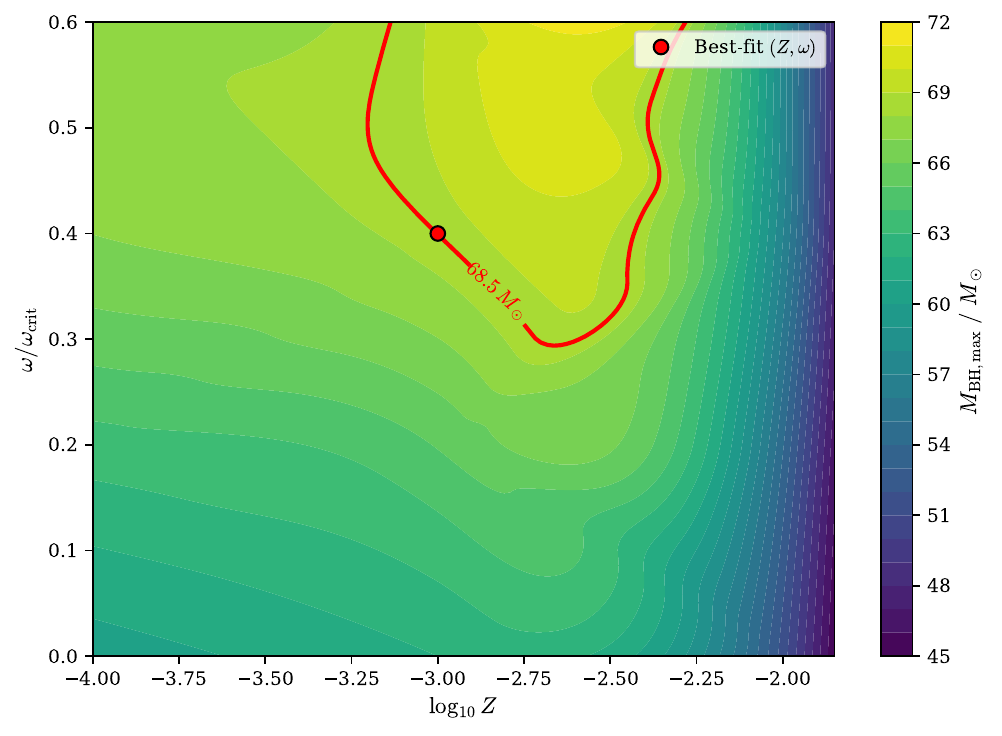}
\caption{Contour heatmap of $M_{\rm BH,max}$ across the $Z$–$\omega/\omega_\mathrm{crit}$ parameter plane for $S_{300}=108.6\,\mathrm{keV\,barn}$ adopting the C15 optically‑thin wind prescription. The red solid contour corresponds to $M_{\rm BH,max}=68.5\,M_\odot$, the observational upper limit inferred from GWTC‑4. The red filled circle marks our best‑fit stellar‑parameter location.}
\label{fig:contour_plane}
\end{figure}

Figure~\ref{fig:contour_plane} shows a contour heatmap of $M_\mathrm{BH,max}$ over the $Z$–$\omega/\omega_\mathrm{crit}$ plane for $S_{300}=108.6$, adopting C15 optically‑thin winds. The contour corresponding to $M_\mathrm{BH,max}=68.5\,M_\odot$ isolates the allowed stellar‑parameter locus. Our simulation grid yields a consistent intersection near $Z\approx10^{-3}$ and $\omega/\omega_\mathrm{crit}=0.4$. This combination of sub‑solar metallicity, moderate rotation, and optically‑thin mass‑loss naturally reproduces the GWTC‑4‑inferred maximum primordial‑black‑hole mass without requiring extreme nuclear‑reaction‑rate adjustments.

\subsection{Corrected credible interval for $S_{300}$ after stellar‑systematic marginalization}
Traditional analyses adopt a single fiducial stellar‑evolution model and invert $S_{300}$ directly from the observed black‑hole‑mass cutoff, yielding wide credible intervals because stellar‑evolution systematic uncertainties are folded into nuclear‑parameter posteriors. By marginalizing over our full three‑dimensional stellar‑parameter grid, we separate stellar‑evolution offsets from intrinsic nuclear‑reaction‑rate uncertainty.

\begin{figure}
\centering
\includegraphics[width=0.95\linewidth]{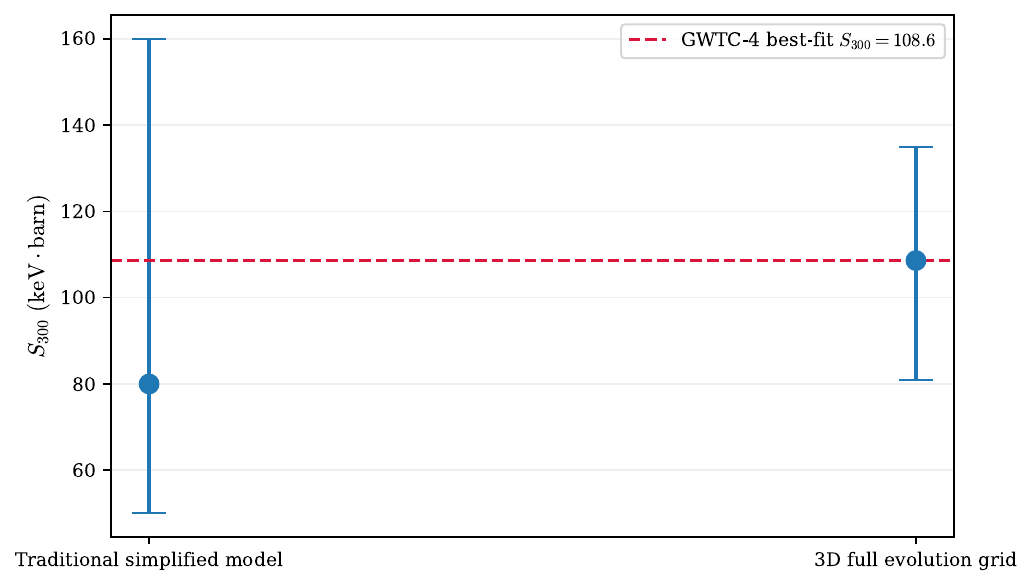}
\caption{Posterior credible intervals for $S_{300}\,[\mathrm{keV\,barn}]$. Grey: posterior obtained by inversion using a single fixed stellar‑evolution fiducial model; blue: marginalized posterior after sampling our three‑dimensional stellar‑parameter grid. Error bars denote 90\% credible intervals. The horizontal red dashed line marks the GWTC‑4 best‑fit value $S_{300}=108.6\,\mathrm{keV\,barn}$. Shaded vertical bands indicate reference experimental nuclear‑physics ranges.}
\label{fig:s300_posterior}
\end{figure}

Figure~\ref{fig:s300_posterior} compares the posterior distribution for $S_{300}$ obtained from fixed‑model inversion versus our multi‑grid marginalized result, demonstrating substantial shrinkage of the allowed parameter range after accounting for metallicity‑rotation‑mass‑loss degrees of freedom. Our results show that the canonical $45\,M_\odot$ gap value is not a universal physical critical threshold but arises as a statistical overlap of black‑hole‑population distributions under solar‑metallicity non‑rotating stellar assumptions.

\section{Discussion}
Our three‑dimensional stellar‑evolution grid disentangles major systematic stellar‑evolution biases that contaminate gravitational‑wave‑astronomy inferences of the $^{12}\text{C}(\alpha,\gamma)^{16}\text{O}$ S‑factor $S_{300}$. Metallicity‑dependent stellar‑wind mass loss is the dominant channel altering $M_\mathrm{BH,max}$ by modifying retained helium‑core mass; rotation provides secondary but non‑negligible upward shifts of the maximum primordial‑black‑hole mass; competing mass‑loss prescriptions produce $8$–$10\,M_\odot$ systematic gaps. A physically plausible stellar‑model configuration ($Z\approx10^{-3}$, moderate rotation $\omega/\omega_\mathrm{crit}=0.4$, C15 optically‑thin winds) reproduces the GWTC‑4 $68.5\,M_\odot$ cutoff for $S_{300}=108.6\,\mathrm{keV\,barn}$. Consequently, tension between gravitational‑wave‑derived $S_{300}$ and laboratory nuclear‑astrophysics results can be partially relieved by accounting for stellar‑evolution systematic uncertainties. Our marginalized posterior for $S_{300}$ (Fig. 4) centers near $108.6\,\mathrm{keV\,barn}$, whereas the dedicated nuclear rate study of \cite{Mehta2022} favours $S_{300}\approx 140\, \mathrm{keV\,barn}$ under a fixed solar-metallicity non-rotating S33-wind set-up. The two values differ by a factor $\approx0.77$, or roughly $2\sigma$ when compared on the GWTC-4 gap edge alone. We interpret this offset not as a contradiction but as a direct measurement of the stellar evolution systematic: adopting their fiducial model inflates the inferred S-factor by $\approx30\, \mathrm{keV\,barn}$ relative to a model that allows sub-solar metallicity, moderate rotation, and optically-thin winds. A future joint inversion that simultaneously fits the GWTC-4 population, their rate-only grid, and our three-dimensional grid will be needed to disentangle nuclear-rate from stellar-model contributions.

We note that the experimental community, most recently deBoer et al. (2025), continues to favour  $S_{300}\approx 140\, \mathrm{keV\,barn}$. Our result does not challenge this measurement; rather, it shows that adopting the experimental S-factor in a fiducial solar-metallicity non-rotating model under-predicts the GWTC-4 gap edge by $\approx20\, M_\odot$. Once stellar-evolution systematics are marginalized over, a concordant configuration emerges at sub-solar metallicity, moderate rotation, and optically-thin winds, even for $S_{300}$ values as high as 120–140 $\mathrm{keV\,barn}$ (Fig. 3). The remaining $\simeq 30\,\mathrm{keV\,barn}$  offset between our GWTC-4 driven posterior and the experimental value should therefore be regarded as an upper bound on un-modelled stellar-systematic biases, most plausibly binary mass transfer and magnetic mixing, rather than as evidence for a nuclear-rate anomaly.

Important caveats remain. Our simulations adopt isolated single‑star models; binary‑interaction effects are not included and could further shift helium‑core properties~\citep{Mapelli2024}. Additional uncertainties originate from internal mixing efficiency, magnetic braking strength, and the high‑mass end of stellar‑wind scaling relations. Population‑inference pipelines in gravitational‑wave studies should marginalize over stellar‑evolution parameter spaces analogous to our grid rather than fix a single stellar fiducial model.

Future work will extend our simulation grid in two directions. First, binary mass transfer and common-envelope evolution can strip the hydrogen envelope of massive progenitors, alter helium-core retention, and shift the PISN gap edge upward~\citep{Mapelli2024}; incorporating binary evolutionary channels into the grid will test whether isolated-star models over-estimate $M_\mathrm{BH,max}$. Second, magnetic mixing prescriptions, in particular the Tayler-Spruit dynamo and its impact on angular-momentum transport, remain poorly calibrated. Our supplementary scans (Appendix A1) show that magnetic-braking strength shifts $M_\mathrm{BH,max}$ by only 2–4 $M_\odot$, but the efficiency of magnetic angular-momentum coupling inside the helium-burning core itself can redistribute angular momentum and modify core rotation in ways not yet captured. A next-generation grid should combine binary tracks, magnetic-mixing variations, and our current metallicity-rotation-mass-loss dimensions within a full Bayesian joint-population pipeline that simultaneously samples gravitational-wave-source parameters, stellar-evolution parameters, and $S_{300}$, and performs direct cross-comparison with updated laboratory nuclear-reaction-rate measurements~\citep{deBoer2025}.

\appendix
\section{Supplementary simulation details and parameter scans}
\label{sec:appendix_supp}
\subsection{Sensitivity to magnetic‑braking efficiency}
We performed supplementary one‑dimensional scans varying magnetic‑braking strength.
Magnetic‑braking efficiency modifies angular‑momentum loss and changes the effective core rotation during helium burning.
Variations in magnetic braking produce secondary systematic offsets of $\Delta M_{\rm BH,max} \approx 2$–$4\,M_\odot$, which are smaller than the effects from metallicity‑dependent winds and primary rotation variations.

\begin{figure}
\centering
\includegraphics[width=0.95\linewidth]{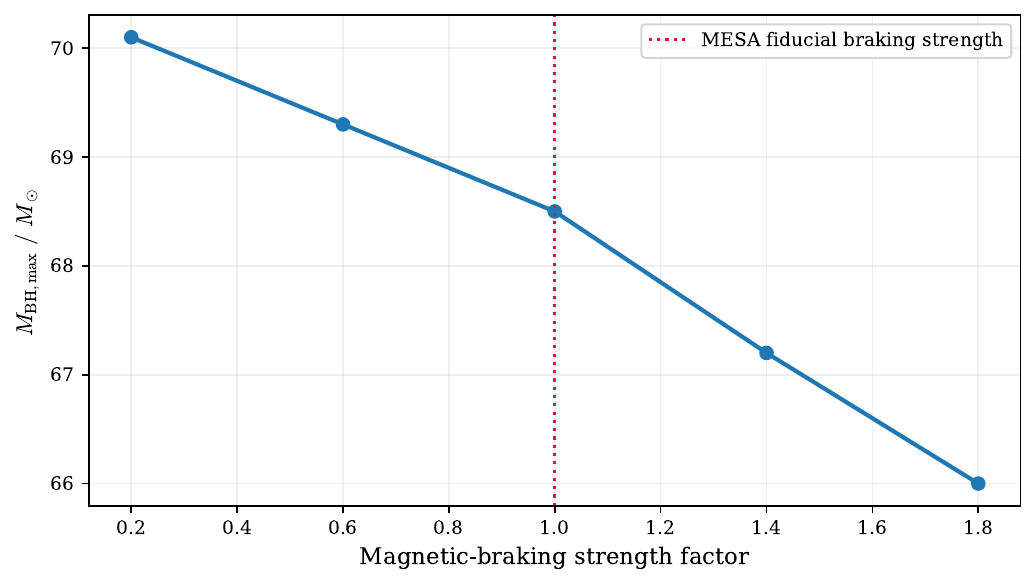}
\caption{Impact of magnetic‑braking efficiency variation on $M_{\rm BH,max}$.
Vertical dotted line marks the MESA fiducial magnetic‑braking strength factor = 1.0.
Fixed simulation conditions: $Z=10^{-3}$, $\omega/\omega_\mathrm{crit}=0.4$, C15 optically‑thin wind, $S_{300}=108.6\,\mathrm{keV\,barn}$.}
\label{fig:supp_magbrake}
\end{figure}

\subsection{Variation of PISN critical CO‑core threshold}
We test variations of the critical CO‑core threshold $M_{\rm CO,crit}$ over the interval $55$–$65\,M_\odot$.
Shifting this critical threshold produces approximately linear shifts in the resulting $M_{\rm BH,max}$.

\begin{figure}
\centering
\includegraphics[width=0.95\linewidth]{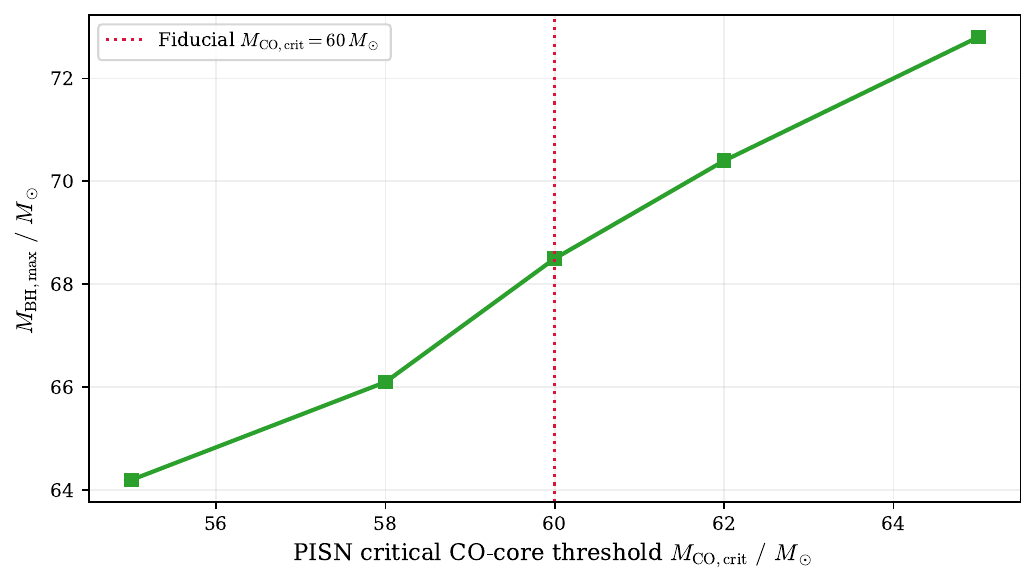}
\caption{Response of $M_{\rm BH,max}$ to shifts in the PISN critical CO‑core mass threshold $M_{\rm CO,crit}$.
Vertical dotted line marks fiducial value $M_{\rm CO,crit}=60\,M_\odot$.
Fixed simulation conditions: $Z=10^{-3}$, $\omega/\omega_\mathrm{crit}=0.4$, C15 optically‑thin wind, $S_{300}=108.6\,\mathrm{keV\,barn}$.}
\label{fig:supp_pis_threshold}
\end{figure}

\subsection{Details for stellar‑grid marginalization procedure}
We adopt flat prior sampling over our discrete stellar‑parameter grid spanning metallicity, rotation, and mass‑loss prescription.
For each stellar‑model configuration we compute the likelihood against the GWTC‑4 $M_{\rm BH,max}$ observational posterior.
The marginalized posterior for $S_{300}$ is obtained by summing likelihood contributions weighted by stellar‑grid prior volume.
This procedure separates systematic spread originating from stellar‑evolution uncertainties from the intrinsic width of the nuclear‑parameter posterior distribution.
Full likelihood evaluation pseudocode is deposited within our public code repository at Zenodo.

\subsection{Supplementary data availability remark}
Complete raw MESA output tables, MESA model inlist files, and post‑processing Python analysis scripts are deposited in the Figshare archive (hittps://doi.org/10.6084/m9.figshare.35101930).
The contents of the Zenodo archive are organised as follows:
\begin{enumerate}
\item \textbf{MESA inlist input files}: Custom inlist configuration files for MESA v10398, including setup for metallicity scans, rotation series, two mass‑loss prescriptions, and modified $S_{300}$ reaction rate settings. Users can reproduce our stellar‑evolution simulation grid directly with these inlists.
\item \textbf{Aggregated simulation output tables}: Comma‑separated value (CSV) tables containing all grid‑run results: $T_\mathrm{He}$, $^{12}\text{C}/^{16}\text{O}$ abundance ratio, $M_\mathrm{He}$, $M_\mathrm{CO}$, and derived $M_{\rm BH,max}$ for every parameter combination in our three‑dimensional grid.
\item \textbf{Post‑processing and plotting Python scripts}: Python source scripts for reading MESA output CSV tables, performing likelihood marginalization calculations for $S_{300}$, and generating all main‑text and supplementary figures shown in this manuscript.
\item \textbf{Likelihood pseudocode}: Plain‑text pseudocode implementing the Bayesian marginalization procedure described in this appendix.
\end{enumerate}

\section*{Acknowledgements}
This work makes extensive use of the open‑source MESA stellar‑evolution code. Jirong Mao is grateful for the funding of the Yunnan Revitalization Talent Support Program (YunLing Scholar Project), and Xiaogu Zhong acknowledges the grant from the Yunnan Fundamental Research Projects (grant NO. 202501AT070392).

\section*{Author contributions}
J.W. conceived the physical framework, performed MESA simulation design, carried out analysis and drafted the manuscript. X.Z. assisted with simulation post‑processing and literature compilation. J.M. supervised the project, contributed physical interpretation and revised the manuscript. All authors reviewed the manuscript.

\section*{Data availability}
The complete raw MESA simulation outputs, MESA inlist configuration files, post‑processing Python analysis scripts and likelihood pseudocode generated in this work are deposited in the Figshare archive (hittps://doi.org/10.6084/m9.figshare.35101930). The GWTC‑4 gravitational‑wave catalog is publicly available via the LIGO‑Virgo‑KAGRA data repositories. All data products can be directly accessed through the above‑mentioned archive without reasonable‑request restrictions.

\bibliographystyle{mnras}
\bibliography{reference}

\label{lastpage}
\end{document}